# A DUAL BAND PRINTED F-ANTENNA USING A TRAP WITH SMALL BAND SEPARATION

Prasad Samudrala [1], Justin Jose [2], and [3]Amit Kulkarni, Ph.D.

Honeywell Building Technologies, Advanced Technology Group, Honeywell Inc.

(1) prasad.samudrala@honeywell.com   (2) justin.jose@honeywell.com   (3) amit.b.kulkarni@honeywell.com

*ABSTRACT*

*Trap antenna is well known method and has many applications. With this method, trap(s) are used on antenna to block currents of some frequencies and so electrically divide the antenna into multiple segments and thus one antenna can work on multiple frequencies. In this paper, trap antenna method is used to design a dual band Sub GHz printed F-antenna. The antenna is printed on FR4 board to achieve low cost solution. The two bands are 865-870 MHz and 902-928 MHz. The challenge of this design is that the frequency separation of the two bands is very small. In this case, and also the extra section for low frequency band is too small. Then, the influence of trap LC component variation due to tolerance to the two resonant frequencies is big, and so it is difficult to achieve good in band return loss within the LC tolerance. This is the main difficulty of this design. The problem is solved by placing the low band section away from the end of the antenna.*

## 1. INTRODUCTION

Adding traps on antenna is a commonly used method to make multiband antennas. In [1], this method was used to design Dual-Band PIFA. In [2], traps were added to wires to make Dual-Band Quadrifilar Helix Antenna. In [3], the trap method was used not just to design multiband antenna, but to design broad band antenna.

In this paper, the trap method is used to design a dual band printed F-Antennas. The design consists of an inverted F antenna, with a trap for creating the dual bands. The two bands are 865-870 MHz and 902-928 MHz. The design goal is that, the return loss for the two bands is under -10dB within the trap LC component tolerance.

For trap F-antenna, a straight forward configuration of the antenna is shown in Fig. 1. A), where the low band section is at the end of the antenna. But this configuration will not work for this antenna design, because the separation of the two bands is small, and thus that section is small. And because this section is small, the influence of trap L and C components variation due to tolerance to the two resonant frequencies is big. Actually, it is too big. It is not possible to achieve the design goal, which is, the return loss for the two bands is under -10dB within trap LC component tolerance.

To solve this problem, the low band section is placed away from the end of the antenna as shown in Fig. 1. B), so it becomes bigger. When it is bigger, the LC tolerance's influence is smaller, then the design goal is achieved.

The first part of the paper describes the design of the trap, second part will present the inverted F-Antenna design. The antenna is designed to be printed on FR-4 material to minimize space and cost of the board. For all simulations, the software used in the designs is CST (Computer Simulation Technology).

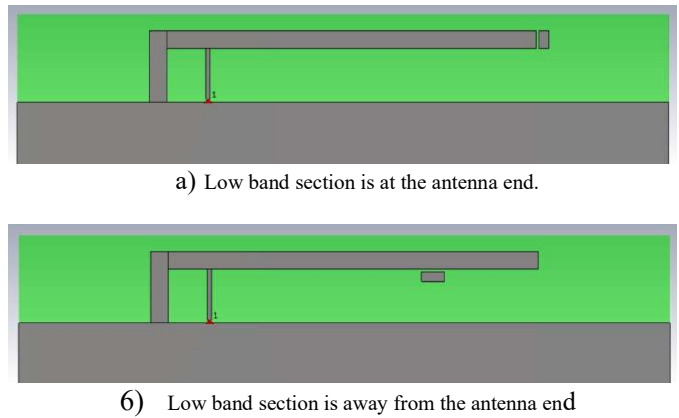

a) Low band section is at the antenna end.

6) Low band section is away from the antenna end

Figure 1. Two congurations of trap antenna

## 2. THE TRAP DESIGN

The Trap used in the design are as follows. It consists of one capacitor and two inductors connected in parallel which acts as a band stop filter for high band. The reason of using two inductors in parallel is for getting the required value.

LC selection: the readily available tolerance for C is +-0.05pF but up to the value 9.1pF in a popular source. For the percentage reason, the highest value 9.1pF is selected. The L value is selected to make the LC trap resonating at 915 MHz. The readily available tolerance for L is +-2%.

The selected capacitor and inductor nominal values and tolerances are C=9.1pF+-0.05pF, L=6.8nF+-0.02%. The trap S21 with the nominal values, and the two ends of tolerances are shown in Fig. 2. Trap S21 variation due to tolerance is clearly seen.

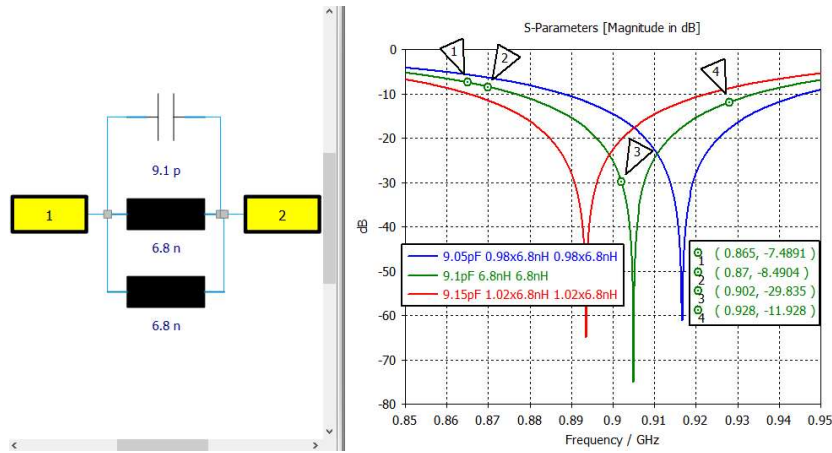

Figure 2 . Trap used in the design

## 3 . THE INVERTED F-ANTENNAS DESIGN

The inverted F antenna design, as shown in Fig. 3, combines a single inverted F antenna with a trap. The ground plane is modeled as 1.6 mm thick,100x100mm PEC. The antenna area is 13.5 mm wide FR4 (Er 4.3, LossTan 0.025). The antenna dimensions are 59.4x11 mm, antenna trace is 2.7 mm wide, feed width is 0.65 mm, antenna PEC thickness is 0.035 mm. Fig. 5 shows the antenna return loss with trap LC nominal values, and the two ends of tolerances. As can be seen, the design goal is achieved, which is, under -10dB for all cases. Fig. 6 shows the antenna 3D far field pattern, gain and efficiencies of the two frequencies with LC nominal values.

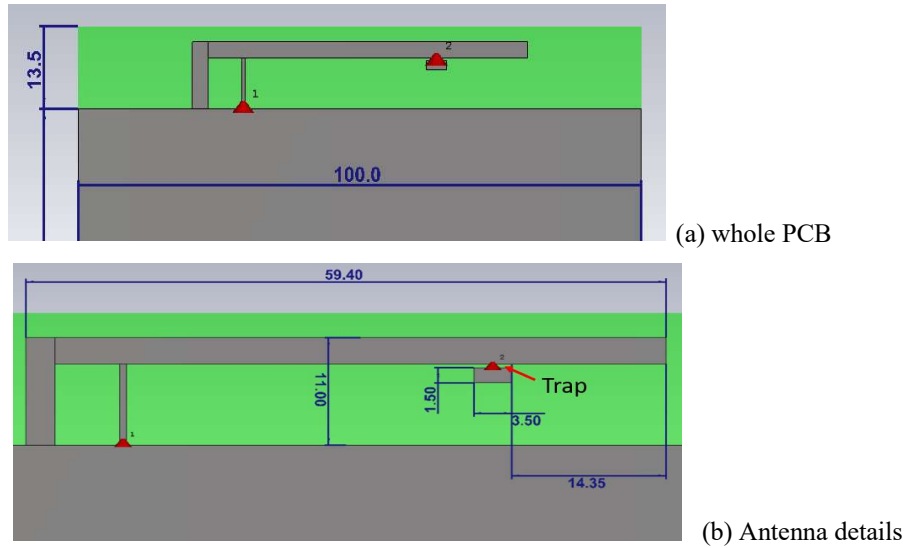

(a) whole PCB

(b) Antenna details

Figure 3 . The design. (a) whole PCB (b) Antenna details

## 6. RESULTS & ANALYSIS

This section deals with the results of the propsoed Dual band Sub GHz antenna . It mainly includes the Surface current distribution, Return Loss ( dB ) , 3D Radiation pattern & Antenna efficiency ( dB ).

As in Figure 4 , the surface current distribution shows strong current distribution along the antenna traces , at both the SubGHz bands , 866 MHz & 915 MHz respectively

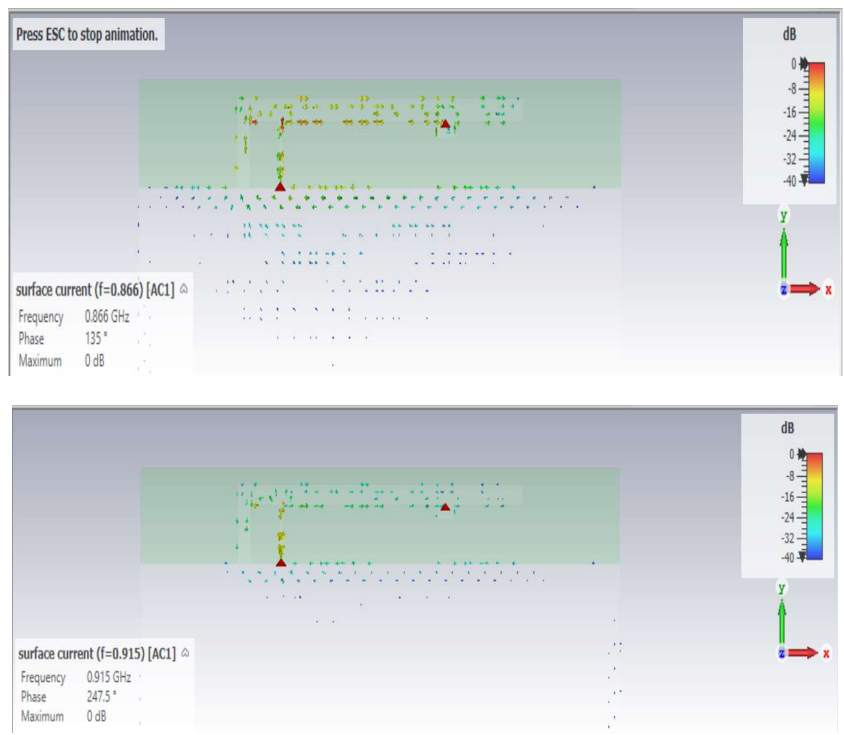

Figure 4 . Surface current distribution at 866 MHz & 915 MHz

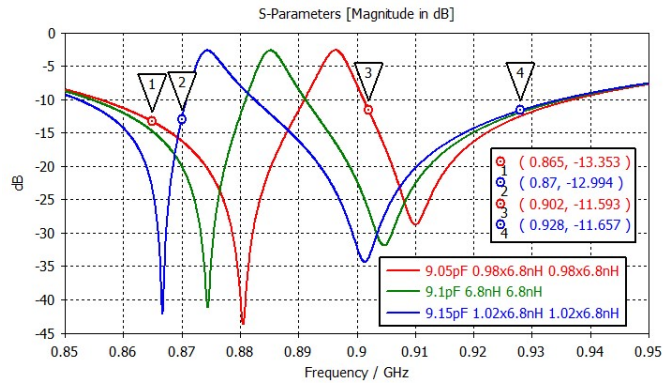

Figure 5. Antenna return loss

The Antenna return loss plot in Figure 5., shows the bandwidth numbers, which is more than 20 MHz for the 868 MHz band (Europe ) & approx.. 50 MHz for 915 MHz which is well beyond the bandwidth requirements .

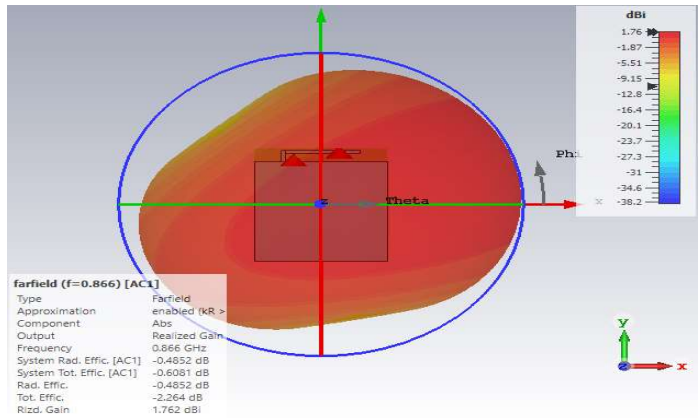

@ 868 MHz Efficiency : -0.61 dB

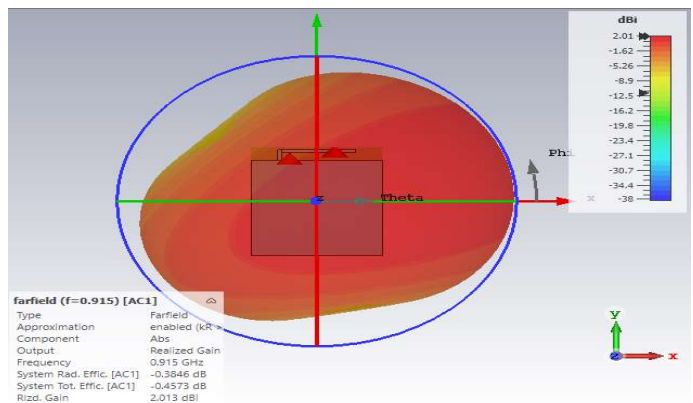

@ 915 MHz efficiency : -0.46 dB

Figure 6. 3D Radiation Pattern and antenna efficiency

The Radiation patterns shown in Figure 6 ., states that the antenna radiates a quasi omni directional pattern and having an efficiency of above 87 % for both the Sub GHz bands .

## 5 . CONCLUSION

In this paper, inverted F-antenna with trap is used to design dual band (860-870 MHz & 902-928 MHz) F-antenna. The antenna is printed on FR4 board for low cost. Normally the low band section of a trap antenna is placed at the end of the antenna. But it does not work when the two bands separation is small due to LC value tolerance. The problem is solved by placing the low band section away from the end of the antenna. A design is done with good result.

### Authors

| | |
|---|---|
| **Prasad Samudrala** has master's degree in Information and communications engineering from Anna University, Chennai. He has over 19 years of experience in Industrial and building automation, IoT, Wireless and Embedded product development. He has over 25 patents and 3 trade secrets to his name. Currently he is in a Principal System Architect position in Honeywell building automation division. His area of interest include Low power wireless technologies, 5G. | 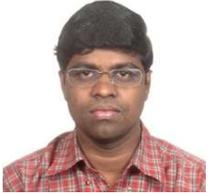 |
| **Justin Jose** is Sr. RF Engineer in Honeywell Building Technologies group . He has more than 8 years of Industrial experience and holds Master Degree in Electronics & Communication Engineering , specialized in Microwave & Antennas from Cochin University of Science & Technology , India . Main areas of interest include Antenna , RF & Wireless Communication and worked in Fixed Wireless , Automotive and Consumer Electronics domains . He has got 2 patents filed and continuing the R& D on various existing & futuristic Wireless technologies . | 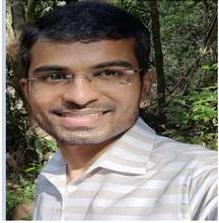 |
| **Amit Kulkarni** is an Engineering Director in HBT Architecture and Innovation team. He also leads the Wireless COE. Amit joined Honeywell in October 2008. Before joining Honeywell, he worked in General Electric Global Research as a Senior Scientist and then became Sr Engineering Manager at GE Security. Amit has 23 years of experience in Wireless Communication Systems. He holds a Ph.D in Computer Science from the University of Kansas. He is an inventor on more than 25 patents and has published 17 papers in various journals and conferences. | 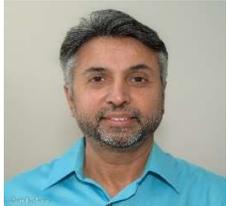 |